# Energy Efficient Automated Driving as a GNEP: Vehicle-in-the-loop Experiments

Viranjan Bhattacharyya, Tyler Ard, Rongyao Wang, Ardalan Vahidi, Yunyi Jia, and Jihun Han

*Abstract*—In this paper, a multi-agent motion planning problem is studied aiming to minimize energy consumption of connected automated vehicles (CAVs) in lane change scenarios. We model this interactive motion planning as a generalized Nash equilibrium problem and formalize how vehicle-to-vehicle intention sharing enables solution of the game between multiple CAVs as an optimal control problem for each agent, to arrive at a generalized Nash equilibrium. The method is implemented via model predictive control (MPC) and compared with an advanced baseline MPC which utilizes unilateral predictions of other agents' future states. A ROS-based in-the-loop testbed is developed: the method is first evaluated in software-in-the-loop and then vehicle-in-the-loop experiments are conducted. Experimental results demonstrate energy and travel time benefits of the presented method in interactive lane change maneuvers.

*Index Terms*—Connected automated vehicles, optimal control, game theory, model predictive control, vehicle-in-the-loop experiments.

## I. Introduction

CONNECTED and automated vehicles and their energy benefits [1], [2] have been a focus of automotive research in recent years as vehicle connectivity is becoming increasingly prevalent in new cars [3] along with advancements in vehicle autonomy [4]. Substantial CAV research has been focused on longitudinal only control, primarily owing to convexity of the problem and the direct relation between longitudinal velocity and acceleration with energy consumption [5]. However, most urban roads and highways in the US have at least 2 lanes and often involve impediments. Hence, in many practical scenarios, car-following behavior is not sufficient and motion planning for energy efficiency needs to incorporate the lateral degree of freedom.

From a predictive control perspective, CAVs could solve the problem of predicting the future intentions of neighboring interacting vehicles by sharing their plans via vehicle-to-vehicle (V2V) communication. However, the plans are interdependent and lane changing introduces richer interactions among CAVs as they can affect each other's costs through additional maneuvers. In particular, we consider a problem where CAVs need to maneuver around an impeding vehicle on the right lane. Such situations arise in congestion caused by slowing buses and other commercial vehicles [6] on urban roads, construction vehicles, vehicle breakdowns or simply due to varying driving speed preference [7]. We focus on energy efficient motion planning in such interaction intensive situations and work at the intersection of game theory and optimal control. Furthermore, we seek a distributed control approach for practical implementation. Therefore, the modeling unit is the individual agent (CAV) and we take the non-cooperative game theory route. We highlight that "non-cooperative" game does not necessarily mean adversarial agent interaction. It merely means that the individual agent is the modeling unit rather than groups of agents as in cooperative game theory. A detailed discussion on these two approaches can be found in [8]. We aim to solve for the Nash equilibrium since we treat each agent equally and seek solutions where no agent can unilaterally perform better. In classical non-cooperative game theory, the solution concept of Nash equilibrium is based on cost (payoff) minimization (maximization) only. However, multi-vehicle motion planning needs incorporation of individual dynamics and coupled collision avoidance constraints. Mathematically, this means that the action set of each agent depends on actions of other agents. In game theoretic terms, this constrained Nash equilibrium problem is a generalized Nash equilibrium problem (GNEP) [9].

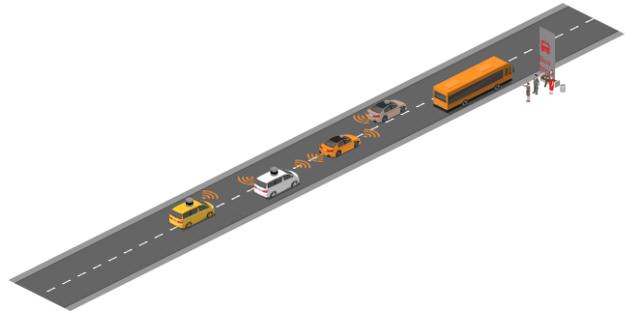

Fig. 1: Scenario illustration: Four CAVs communicate to negotiate their maneuvers around a slowing bus.

GNEP for interactive motion planning in automated driving has been studied in recent works such as [10]–[12]. In [10] an optimal control solution to GNEP is analyzed along with simplified abstract simulations in a multi-vehicle scenario. GNEP where each agent's cost function depends only on its states and controls have been shown to be solvable as an optimal control problem (OCP) to arrive at a Nash equilibrium, with convexity assumptions on control set and dynamics, and compactness of state sets. Distributed solution of GNEPs as

Viranjan Bhattacharyya and Ardalan Vahidi are with the Dept. of Mechanical Engineering, Clemson University, SC, 29634 USA {vbhatta, avahidi}@clemson.edu

Rongyao Wang and Yunyi Jia are with Dept. of Automotive Engineering, Clemson University, SC 29607, USA {rongyaw, yunyij}@clemson.edu

Tyler Ard and Jihun Han are with the Vehicle and Mobility Systems Dept. at Argonne National Laboratory, Lemont, IL 60439, USA {tard, jihun.han}@anl.gov

potential games was discussed in [13]. It has been studied in the multi-vehicle automated driving planning context in [12] and iterative algorithms have been discussed. Often, driving decision making, such as lane choice and relative spatial positioning, lends itself well to the mixed logical dynamical (MLD) framework [14] where logical decisions are optimized via integer variables which are integrated with vehicle motion models and constraints. The MLD framework has been leveraged in [15]–[17] for modeling the inter-vehicle interactions in lane change and merge scenarios. A multi-agent CAV planning schema has been discussed in [16] through simulations. Toward an equilibrium solution, an $\epsilon$-mixed integer Nash equilibrium has been defined and iterative solution methods have been presented in [11] which are validated via simulations.

*Contributions*

The key contributions of this work are summarized as following:
- Formulation of multi-agent energy efficient automated driving motion planning in interactive lane change situations as a GNEP. Analysis formalizing V2V intention-sharing's facilitation of solution of the GNEP as an OCP for each agent is presented. This enables a distributed model predictive control implementation.
- Experimental platform development and testing of the proposed method in software and vehicle-in-the-loop experiments. The proposed method is compared with an advanced baseline MPC which utilizes a unilateral prediction model for predicting future states of other vehicles.

The paper is structured as follows: in Section II the problem is formulated and the solution concept is presented. In Section III, the experiments are described, which includes development of a baseline predictive model, testbed description, software and vehicle-in-the-loop tests and results. Finally, in Section IV we conclude.

## II. PROBLEM FORMULATION

The problem of multiple automated vehicles (agents) attempting to maneuver around an impeding vehicle while minimizing a cost can be formulated as a GNEP. This is because the collision avoidance constraints for each agent are a function of states of the other agents in the game. Keeping the discussion reasonably self-contained, we provide definition of a GNEP in the controls context.

**Definition 1.** *A game between agents $v \in \{1, ..., N\}$ is a GNEP when each agent with states $x^v$ and controls $u^v$ solves the problem*

$$\min_{u^v} \quad J^v(x^v, u^v, x^{\neg v}, u^{\neg v}) \tag{1a}$$
$$\text{s.t.} \quad x^v \in \chi_v(x^{\neg v}) \tag{1b}$$
$$u^v \in \mathrm{U} \tag{1c}$$

*where, the constraint set (1b) of an agent $v$ is a function of states of other agents $\neg v$.*

It may be noted that in the controls context, since the agents have dynamics, the states are a function of the controls and hence, control is the decision variable for each agent. Predictive control-based planning has been shown to improve energy efficiency of automated vehicles compared to reactive control-based planning [16], [18]. We work at the synergistic intersection of predictive control and algorithmic game theory [19] and formulate the particular problem considered in this work as a multi-stage dynamic GNEP for each interacting automated vehicle $v$ as

$$\min_{u^v} \quad J^v = \sum_{k=0}^{T-1} \{q_{1,k}(\dot{s}_k^v - \dot{s}_{ref}^v)^2 + q_{2,k}[(a_k^v)^2 + (u_{a,k}^v)^2]$$
$$+ q_{3,k}[(l_k^v - l_{ref})^2 + (u_{l,k}^v - l_{ref})^2]\} + q_4(s_T^v - s_{ref}^v)^2$$
$$+ q_{1,T}(\dot{s}_T^v - \dot{s}_{ref}^v)^2 + q_{2,T}(a_T^v)^2 + q_{3,T}(l_T^v - l_{ref})^2 \tag{2a}$$
$$\text{s.t.} \quad x_{k+1}^v = Ax_k^v + Bu_k^v \tag{2b}$$
$$x_k^v \in \mathrm{X}_{safe}(x_k^{\neg v}) \tag{2c}$$
$$u_k^v \in \mathrm{U}_{admissible} \tag{2d}$$

The linear dynamic model from [18] is utilized such that the states of agent $v$ are $x^v = [s^v \ \dot{s}^v \ a^v \ l^v \ \dot{l}^v]^T$: longitudinal position, velocity, acceleration, lane position and lane change rate, respectively while the controls are $u^v = [u_a^v \ u_l^v]^T$: acceleration command and lane number command, respectively, at each time step $k$. The cost function is designed to track a desired longitudinal speed $\dot{s}_{ref}^v$, terminal position $s_{ref}^v$ and reference lane $l_{ref}$, while minimizing the acceleration and acceleration command. This cost function captures travel time preference of each vehicle via velocity tracking and energy consumption reduction via acceleration minimization.

### A. Solution Approach

In general, finding a solution of (2), called the generalized Nash equilibrium (GNE), is challenging due to the coupling in collision avoidance constraints [20], [21]. However, if the current and future states (intentions) of other vehicles are obtainable, then solving the dynamic GNEP (2) reduces to solving an OCP by each agent. Furthermore, since the cost function of each agent in our problem formulation is a function of its own states only, the OCP solution is a GNE (see Theorem 1 in [10]).

**Lemma 1.** *If the optimal states of neighboring interacting agents over-the-horizon, $x^{*\neg v}$, are available then dynamic GNEP (2) can be solved as an optimal control problem for $v$.*

*Proof.* If $x^{\neg v} = x^{*\neg v}$ are available for $k = 0, ..., T$, then $\mathrm{X}_{safe}(x_{k=0,...,T}^{\neg v}) = \mathrm{X}_{safe}$, i.e., the safe set is fully determined over $T$ and (2c) reduces to inequality constraints on $x^v$ only. □

V2V connectivity enables this as all the vehicles can share their intentions with each other. Each vehicle $v$ computes its solution over-the-horizon $T$ by solving (2) with $x_{k=0,...,T}^{\neg v} = x_{k=0,...,T}^{*\neg v}$ available from V2V communication.

We denote the dynamic GNEP (2) of each agent as $G^v(x_0^v, x^{\neg v})$ and the corresponding OCP as $G^v(x_0^v, x^{*\neg v})$ -

where $x_0^v$ is a (feasible) initial condition of the agent and $x^{*\neg v}$ are the (optimal) intentions of others.

**Definition 2.** *The Nikaido-Isoda function [9] for the GNEP (2) can be defined at each time step as*

$$\Psi(u, \tilde{u}) \doteq \sum_{v=1}^{N} [J^v(x^v, u^v) - J^v(\tilde{x}^v, \tilde{u}^v)] \quad (3)$$

*where, $u = [u^1, ..., u^v, ..., u^N]^T$ and $\tilde{u} = [\tilde{u}^1, ..., \tilde{u}^v, ..., \tilde{u}^N]^T$.*

We drop the time step subscript for readability but highlight that this is applied at each time step $k = 0, ..., T$. The Nikaido-Isoda function captures the change in game cost when the agents unilaterally deviate from $(x^v, u^v)$ to $(\tilde{x}^v, \tilde{u}^v)$ by changing their controls.

**Definition 3.** $u^*$ *is a GNE of $G^v(x_0^v, x^{\neg v})$ iff*

$$\sup_{\tilde{u}} \Psi(u^*, \tilde{u}) = 0 \quad (4)$$

This means that at the GNE, no agent can reduce its and the game's cost by changing only its own controls.

**Theorem 1.** *Solution of the OCP $G^v(x_0^v, x^{*\neg v})$ $\forall v$ is a solution (GNE) of the dynamic GNEP (2), $G^v(x_0^v, x^{\neg v})$.*

*Proof.* If $u^{\neg v} = u^{*\neg v}$ then $x^{\neg v} = x^{*\neg v}$ due to dynamics, and by Lemma 1, $G^v(x_0^v, x^{*\neg v})$ is an OCP. Denoting the improvement in functional $J(\cdot)$ with change in $u$ as $\Delta_u J$, if $\bar{u}^v$ is a solution of the OCP $G^v(x_0^v, x^{*\neg v})$ then $\Delta_{\bar{u}^v} J^v \to 0$. Now, since $u^{\neg v} = u^{*\neg v}$, $\Delta_{u^{*\neg v}} J^{\neg v} \to 0$ $\forall \neg v$. Let $u \doteq [\bar{u}^v, u^{*\neg v}]$, then $\Psi(u^*, u) = \Delta_{\bar{u}^v} J^v + \Sigma_{\neg v} \Delta_{u^{*\neg v}} J^{\neg v} \to 0$ which by Definition 3 implies $u = [\bar{u}^v, u^{*\neg v}]$ is a GNE. $\square$

This enables distributed solution of its OCP by each agent resulting in a GNE. In particular, the OCP solved by each agent needs to capture three objectives, namely speed tracking, acceleration minimization and collision avoidance while maintaining road discipline. The objectives are captured by formulating the OCP as a mixed-integer quadratic program (MIQP) which is implemented via MPC.

The MIQP MPC used to numerically solve the dynamic GNEP (2) is based on [18]. In addition to equation (2), an output relation $\Delta u_{l,k} = u_{l,k} - u_{l,k-1}$ defines the lane change direction signal, and $P$ slack variables are introduced to maintain practical feasibility. The cost function is augmented as,

$$\min_{\mathbf{u}, \epsilon, \Lambda} J^v + \sum_{p=1}^{P} q_{\epsilon, p} \|\epsilon_p\|_\infty \quad (5a)$$

where, $J^v$ is the original cost, $\mathbf{u}$ are the original controls, $\epsilon$ are the slack variables and $\Lambda$ are logical (integer) variables.

To model the temporal lag between commanded acceleration ($u_a^v$) and realized acceleration ($a^v$), introduced by the vehicle powertrain response, a lumped first-order model is employed,

$$\tau \dot{a}^v = u_a^v - a^v \quad (6)$$

where $\tau$ is the fixed lag constant. The MPC optimizes for choosing a lane through an integer decision variable. This is the lane command ($u_l$) and the lateral motion is modeled as a step response to this input. A second-order damped model follows,

$$\ddot{l}^v = K_l \omega_n^2 u_l^v - 2\zeta \omega_n \dot{l}^v - \omega_n^2 l^v \quad (7)$$

where $K_l$ is a scalar lateral input gain, $\omega_n$ is the natural frequency response coefficient, and $\zeta$ is the damping ratio. The model decouples the lateral and longitudinal motion and renders a linear model

$$\frac{d}{dt}\begin{bmatrix} s^v \\ \dot{s}^v \\ a^v \\ l^v \\ \dot{l}^v \end{bmatrix} = \begin{bmatrix} 0 & 1 & 0 & 0 & 0 \\ 0 & 0 & 1 & 0 & 0 \\ 0 & 0 & -\frac{1}{\tau} & 0 & 0 \\ 0 & 0 & 0 & 0 & 1 \\ 0 & 0 & 0 & -\omega_n^2 & -2\zeta\omega_n \end{bmatrix} \begin{bmatrix} s^v \\ \dot{s}^v \\ a^v \\ l^v \\ \dot{l}^v \end{bmatrix} + \begin{bmatrix} 0 & 0 \\ 0 & 0 \\ \frac{1}{\tau} & 0 \\ 0 & 0 \\ 0 & K_l\omega_n^2 \end{bmatrix} \begin{bmatrix} u_a^v \\ u_l^v \end{bmatrix} \quad (8)$$

This continuous time model is discretized with a zero-order hold and sampling time $\Delta t_h = 0.4$ s.

*Admissible Lane Change:* Due to the simplified linear model with lateral and longitudinal decouplement, the optimizer may switch rapidly between discrete lane commands (akin to pulse width modulation). This can result in high lateral jerk and can destabilize the low-level controller. So, until the vehicle converges within a tolerance of $\delta_l$ to the respective center of the lane, the commanded lane is fixed.

$$\Delta u_{l,k}^v + l_k^v - u_{l,k-1}^v \leq 1 + \delta_l \quad (9a)$$
$$u_{l,k-1}^v - \Delta u_{l,k}^v - l_k^v \leq 1 + \delta_l \quad (9b)$$

Also, lane changes below a speed $\underline{v}$, if any, are prevented by enforcing

$$\Delta u_{l,k}^v - \underline{v}^{-1}(\dot{s}_k + \epsilon_1) \leq \gamma \quad (10a)$$
$$-\Delta u_{l,k}^v - \underline{v}^{-1}(\dot{s}_k + \epsilon_1) \leq \gamma \quad (10b)$$

where $\gamma = 1e^{-3}$ is a tolerance. In effect, whenever vehicle velocity is less than $\underline{v}$, $|\Delta u_{l,k}^v| \leq 1$ (with some slack and tolerance) and since $u_l^v$ is an integer, a change is prevented.

*Collision Avoidance Constraints:* $X_{safe}$ for collision avoidance is enforced by optimizing longitudinal position ($s^v$) and lateral position ($l^v$) of ego vehicle $v$, at each prediction step (notation dropped here), so as to avoid a rectangular set $\mathcal{O}^{\neg v}$ (Fig. 2) around each neighboring vehicle (NV), formed by adding ego vehicle dimensions to the NV. $\mathcal{O}^{\neg v}$ is implemented logically by introducing binary variables.

$$[\lambda^{\neg v} = 1] \leftrightarrow \underline{l}^{\neg v} - \delta \leq l^v \leq \overline{l}^{\neg v} + \delta \quad (11a)$$
$$\lambda^{\neg v} \wedge \sigma_f^{\neg v} \leftrightarrow s^v - \epsilon_3 \leq s_-^{\neg v} - d \quad (11b)$$
$$\lambda^{\neg v} \wedge \sigma_r^{\neg v} \leftrightarrow s^v + \epsilon_3 \geq s_+^{\neg v} + d \quad (11c)$$

$\lambda^{\neg v}$ is triggered to be *true* (=1) iff NV $\neg v$ is longitudinally aligned with the ego vehicle. Further, binary variables $\sigma_f^{\neg v}$ and $\sigma_r^{\neg v}$ respectively indicate whether NV is in front or rear of ego vehicle. Here, $\wedge$ denotes logical and, half ego vehicle width $\delta$ is added as a margin to each lateral boundary of the obstacle $\underline{l}^{\neg v}$ and $\overline{l}^{\neg v}$. Obstacle set front position $s_+^{\neg v} = s^{\neg v} + L$ and rear position $s_-^{\neg v} = s^{\neg v} - L$, where $L$ is half of vehicle length. A gap margin $d$ is added in the constraint for safety.

Naturally, NV can only be either in front of *or* behind the ego vehicle, so the following constraint is imposed.

$$[\lambda^{\neg v} = 1] \implies \sigma_f^{\neg v} + \sigma_r^{\neg v} = 1 \quad (12)$$

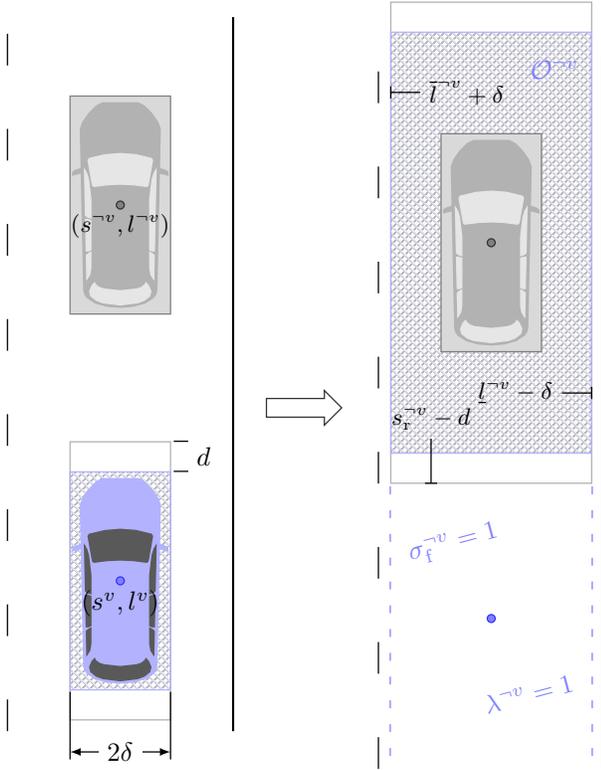

Fig. 2: Example construction of $\mathcal{O}^{\neg v}$ and subsequent activation of avoidance constraints (11) around an obstacle $\neg v$. In this example illustration, $\neg v$ is longitudinally in line with ego $v$ so $\lambda^{\neg v} = 1$ – and in front of ego so $\sigma_{\text{f}}^{\neg v} = 1$ and $\sigma_{\text{r}}^{\neg v} = 0$.

Notably here, the set of logical constraints (11) and (12) form a special ordered set (SOS) which can be exploited when choosing branching rules to improve solution computation speeds [22]. The avoidance constraint regions are thus visualized in Figure 2. Logical activation of the mixed-binary constraints are handled via the Big-M method. An example of the programming details can be found in [23].

The MPC is coded in C++ and the resulting MIQP is solved via Gurobi which utilizes the branch-and-cut algorithm [22]. The algorithmic implementation is referred to as GNEP-MPC and is described in Algorithm 1.

---

**Algorithm 1** GNEP-MPC Planner

For all $v$: initialize $x_0^v$, list $x^{intent}[v] = None$
**while** $s_0^1 \neq s_{end}$ **do**
  **for** $v \in \{1, ..., N\}$ **do**
    get $x_0^v$ from low-level
    **if** $x^{intent}[\neg v] \neq None$ **then**
      $x^{*\neg v} \leftarrow x^{intent}[\neg v]$
    **else** $x^{*\neg v} \leftarrow$ predict($\neg v$)
    **end if**
    $[x_{k=0:T}^{*v}, u_{k=0:T}^{*v}] \leftarrow$ solution of $G^v(x_0^v, x^{*\neg v})$
    $x^{intent}[v] \leftarrow x_{k=0:T}^{*v}$
  **end for**
**end while**

---

During initialization, the NV trajectories are predicted via predict() using the prediction model described in Section III-A.

### B. Interaction

This work focuses on interaction intensive scenarios with the aim of maneuvering CAVs so as to minimize energy consumption while balancing travel time preferences. We define intense interaction by defining a strong dynamic GNEP as follows.

**Definition 4.** *Let $X$ be the state space of agents. The dynamic GNEP $G^v(x_0^v, x^{\neg v})$ is strong if $\{x^v \in X : |x^v - x^{\neg v}| \preceq \mathbf{d}\} \quad \forall v$.*

The vector $\mathbf{d} \in \mathbb{R}^5$ is a parameter which defines agent pairwise compact sets in the state space ($|\cdot|$ is applied elementwise).

**Remark 1.** *In strong interaction scenarios, the agents are able to influence each other's cost substantially.*

Outside of these compact sets, the game becomes trivial as each agent is able to minimize its cost *irrespective* of other agents' actions.

## III. EXPERIMENTS

We first evaluate the presented GNEP-MPC in software-in-the-loop (SiL). This serves as a precursor to vehicle-in-the-loop (ViL) experiments in which one CAV is a retrofitted Mazda CX-7. We focus on a scenario (Figure 1) where four CAVs need to negotiate a lane change around an impeding vehicle on a two-lane road. We compare the GNEP-MPC with a baseline MPC employing a unilateral prediction model for trajectory predictions of NVs. For the baseline, we develop a prediction model of NVs which has an Intelligent Driver Model [24] for longitudinal motion prediction and a constant lateral velocity assumption for lateral motion prediction.

### A. Baseline Prediction Model

We will compare the algorithm proposed here to a strong baseline MPC. In this section we describe this baseline algorithm. The baseline model utilizes separate longitudinal and lateral heuristic elements to predict the motion of surrounding traffic as based on [25]. Because of inherent modeling uncertainty, the use of chance constraints to adjust the safety gap margin from the preceding vehicle is additionally utilized.

The model considers available current position, $s_{\text{nv}}$, and velocity, $v_{\text{nv}}$, measurements of up to two preceding vehicles and up to two trailing vehicles in the current and surrounding lanes (if they exist). The measurements are limited to those within a distance range of $250 \, \text{m}$ to account for sensor hardware limitations.

The baseline longitudinal prediction model leverages constant acceleration kinematic relations on rate of change of $v_{\text{nv}}$ and $s_{\text{nv}}$, and it saturates to prevent reversing or speed-limit-violating prediction. As opposed to the previous approach in [26], the surrounding vehicle current acceleration is not used as the integrating value here - given that it can be subject to measurement noise and can change during the horizon. Instead, the model assumes the future acceleration or braking values, $a_r$, of surrounding traffic to reach a desired speed, $v_0$, given a current estimate of its acceleration $a_{\text{nv}}$ and a band on the acceleration measurement noise $\tilde{a} = 0.35\,\text{m/s}^2$.

$$a_r = \begin{cases} a_0 & a_{\text{nv}} \geq \tilde{a} \text{ and } v_{\text{nv}} < v_r \\ -b_0 & a_{\text{nv}} \leq -\tilde{a} \text{ and } v_{\text{nv}} > 0 \\ 0 & \text{otherwise} \end{cases} \quad (13a)$$

$$\dot{s}_{\text{nv}} = v_{\text{nv}} \quad (13b)$$

$$\dot{v}_{\text{nv}} = a_r \quad (13c)$$

Here, $a_0 = 1.15\,\text{m/s}^2$ is the comfortable acceleration, $b_0 = 2.94\,\text{m/s}^2$ is the comfortable deceleration as calibrated from highway data in [27], and $v_r$ is the desired velocity of ego vehicle.

In addition, a naïve constant acceleration prediction of the trailing vehicle behind the ego vehicle can over-predict its motion trajectory, leading to a (false) indication that a rear-end collision is imminent - which can force the ego vehicle to evasively maneuver away. In reality the trailing vehicle is likely to interact and adjust its trajectory in relation to the ego vehicle. As such, some spatial and queuing interactions for the trailing vehicle are considered by leveraging dynamics from the Intelligent Driver Model (IDM).

$$\dot{s}_{\text{nv}} = v_{\text{nv}} \quad (14a)$$

$$\dot{v}_{\text{nv}} = a_0 \left(1 - \left(\frac{v_{\text{nv}}}{v_0}\right)^\delta - \left(\frac{s^*(v_{\text{nv}}, \Delta v)}{\Delta s}\right)^2\right) \quad (14b)$$

with

$$s^*(v, \Delta v) = s_0 + \max\left\{0,\ Tv + \frac{v\Delta v}{2\sqrt{a_0 b_0}}\right\}$$

and where $T = 1\,\text{s}$ is the desired time headway, $\delta = 4$ is a free-flow velocity tracking exponent, $\Delta s$ is the bumper-to-bumper distance from the preceding vehicle, $\Delta v$ is the speed difference from the preceding vehicle, and $s_0 = 4\,\text{m}$ is the desired vehicle-to-vehicle stand-still gap. Additionally, $v_0$ is assumed to be either the same as $v_r$ if traveling slower than this speed or to be $v_{nv}$ if they are traveling faster than $v_r$. Note here, that the bumper-to-bumper distance is readily available from the last predicted trajectory $X$ of the ego vehicle.

The lateral motion of surrounding traffic is additionally predicted. The baseline lateral prediction model activates when a surrounding vehicle is detected as starting a lane change. This assumes a constant rate-of-change kinematic relation on lane position, $\ell_{nv}$, which is normalized by lane width. Similar to the longitudinal prediction, the lateral model saturates when the surrounding traffic is predicted to have reached its immediate next lane, where $\Delta \ell_{nv}$ is the change in lane since the start of prediction. Here, the assumed rate of change of lane, $\zeta_r$, is non-zero if the surrounding vehicle is measured to change lanes with a rate of $\zeta_0 = v_{nv} \sin(\theta_{\text{nv}})$ of at least $0.2\,\ell/\text{s}$.

$$\zeta_r = \begin{cases} \zeta_0 & \Delta\ell_{\text{nv}} \leq 1 \text{ and } \zeta_0 \geq 0.2 \\ -\zeta_0 & -\Delta\ell_{\text{nv}} \leq 1 \text{ and } \zeta_0 \geq 0.2 \\ 0 & \text{otherwise} \end{cases} \quad (15a)$$

$$\dot{\ell}_{\text{nv}} = \zeta_r \quad (15b)$$

Notably, the model predicts that the surrounding vehicle lane change maneuver completes at least within $\zeta_0^{-1}$ seconds.

The prediction is subject to inherent modeling error, which can cause safety concerns in cases that over-estimate the forward positions of surrounding traffic. To mitigate this, chance constraints are introduced so that the bumper-to-bumper distance from both the immediate preceding and trailing vehicles, $d_{\text{pv}}$ and $d_{\text{tv}}$, are larger than a standstill safety gap $d$ at least with probability $\alpha$.

$$\mathbb{P}(d_{\text{pv}} \geq d) \geq \alpha \quad (16a)$$

$$\mathbb{P}(d_{\text{tv}} \geq d) \geq \alpha \quad (16b)$$

Further description of these chance constraints can be found from previous work in [25], [28]. This model is then utilized by a MIQP MPC planner to generate over-the-horizon trajectory predictions of NVs in the absence of V2V intention sharing.

*B. Software-in-the-loop setup*

We have developed a software environment with three major elements: *simulator*, *planner*, and *tracking controller*.

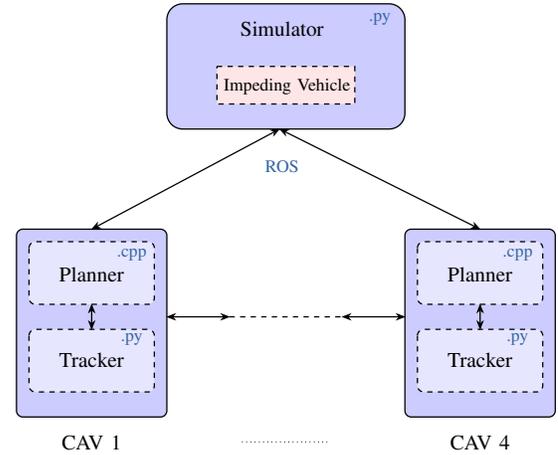

Fig. 3: Architecture

- The simulator handles the road geometry, localization of the vehicles and generates traffic (IDM driven impeding vehicle here) along with visualization.
- The planner is the MIQP MPC which communicates with the low-level tracker for obtaining vehicle states and passing control commands.
- The low-level tracker runs the control commands on a model of vehicle (see Appendix A).

The communications are set up via ROS and each automated vehicle is embedded in the simulation via a dedicated planner and tracking controller communicating with the simulator.

*C. Software-in-the-loop Tests*

We focus on a scenario where four CAVs start longitudinally 15 m apart on the right lane, initially at 0 m/s. The CAVs have different driving speed preferences. An IDM driven impeding vehicle (IV) is spawned at 200 m in the same lane and its maximum velocity is 5 m/s.

CAV 1 is chosen as the global ego vehicle, (it has the maximum number of vehicles ahead) and aims to go from 0 to 600 m. Four sub-scenarios are tested by setting the driving speed preferences as permutations of $(17, 14, 11, 8)$ m/s, respectively, such that the sub-scenarios have strong interaction according to Definition 4 with $\mathbf{d} = [400, 9, \cdot, 2, \cdot]$. The maximum speed is limited to 17 m/s and maximum distance to 600 m due to safety and hardware limitations for eventual ViL experiments. This defines the operational design domain.

*1) Results:* The average improvement over three trials of each sub-scenario are shown in Table I. The energy consumption is calculated using the fuel consumption model presented in [5] as

$$\mathcal{F} = \int_0^{t_f} p_1 u_t v \, dt + \int_{T_1} p_0 \, dt \quad (17)$$

where, $u_t = a + \beta v^2 + C_{rr} g$ is the tractive acceleration accounting for the drag losses $\beta v^2$ and rolling resistance $C_{rr} g$, $v = \dot{s}$, $T_1 = \{t \mid 0 < u_t \leq u_{max}\}$ and $p_0 = 0.371$ g/s and $p_1 = 0.127 \, \text{gs}^2/\text{m}^2$ are fuel model parameters.

TABLE I: Percent improvement in energy consumption and travel time as a result of intention sharing compared with baseline without intention sharing in SiL

| S.No. | Sub-scenario [m/s] | % Energy Ego | % Energy Group | % Travel Time Ego | % Travel Time Group |
|---|---|---|---|---|---|
| 1 | (17, 14, 11, 8) | 8.2 | 12.7 | 7.9 | 6.9 |
| 2 | (11, 14, 17, 8) | 2.9 | 4.3 | 4.3 | 1.7 |
| 3 | (14, 11, 17, 8) | 7.6 | 8.3 | 8.8 | 1.8 |
| 4 | (17, 14, 8, 11) | 3.1 | 7.2 | 7.7 | 0.6 |

From equation (17), it is observed that a trajectory which has low accelerations and maintains a constant longitudinal speed consumes less fuel. The MPC cost has been designed such that vehicles minimize acceleration and track a constant reference velocity. Therefore, a lower MPC cost will cause lower energy consumption. Figure 4 shows the closed-loop cost of ego during three runs in sub-scenario 1. An edge case arose in this scenario where the baseline cost shot up (orange). Overall, with the GNEP-MPC, the closed-loop cost reaches lower values (earlier) compared to the non-game baseline resulting in lower energy consumption. Velocities closer to the reference maintain travel time efficiency.

Physically, the energy savings with GNEP-MPC arise due to reduced braking. This is a consequence of availability of more accurate state predictions of NVs over the horizon in GNEP-MPC compared to the baseline non-game MPC. Harsh corrective actions such as braking and lane changes to avoid collisions are ameliorated. Figure 5 shows comparison of position and velocity profile of the ego vehicle with baseline and GNEP-MPC for the first sub-scenario. In the aforementioned edge case (orange), large braking caused the cost to shoot up with the non-game baseline. Such a situation is mitigated with GNEP-MPC owing to more accurate interactive predictions.

Figure 6 shows the closed-loop trajectories of the NV CAVs along with their V2V shared position predictions over the horizon received by ego at each step in the first sub-scenario. The MPC solutions are open loop GNE. Due to the possibility of deviation in shared (open-loop) and actual (closed-loop) trajectory owing to factors like MPC's dynamics model mismatch, closed-loop GNE convergence cannot be claimed. Nevertheless, receding horizon open loop GNE solutions are able to achieve better control performance in closed-loop as depicted in Figure 4 and summarized in Table I.

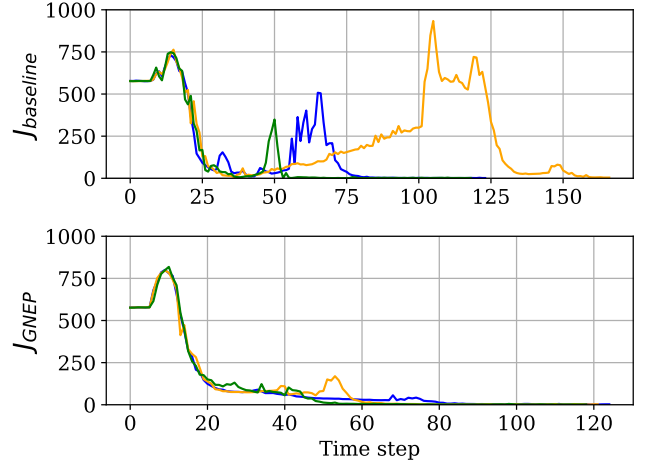

Fig. 4: Closed-loop cost comparison of baseline and GNEP-MPC for the ego vehicle: cost over time during three runs is shown for sub-scenario 1.

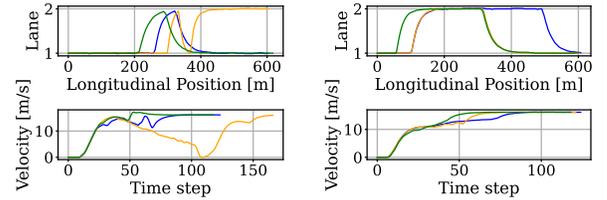

(a) Baseline trajectory  (b) GNEP-MPC trajectory

Fig. 5: Comparison of position and velocity profile of ego vehicle over three runs in sub-scenario 1. GNEP-MPC planning results in reduction of rapid lane change and braking.

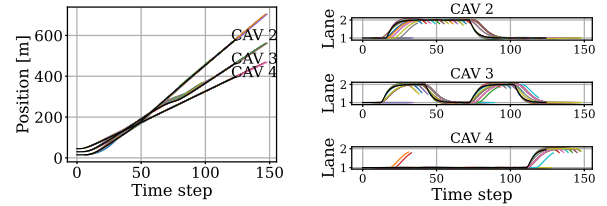

(a) Longitudinal trajectory  (b) Lateral trajectory

Fig. 6: Closed-loop (black) and V2V shared (colored) position of NVs of ego. Shared predictions are shown emanating from the actual trajectory at each time step.

### D. Vehicle-in-the-loop Experiments

The ViL experiments were conducted at the International Transportation Innovation Center (ITIC) test track [29] in Greenville, SC, USA with a retrofitted Mazda CX-7. The SiL architecture has been designed such that any number of automated vehicles, connected or not, can be plugged in. We extend it to ViL simulation by *plugging* in the real vehicle's low-level controller for CAV 1. Thus, a mixed reality environment [30] is created where the simulator simulates a virtual equivalent of the test track with three software CAVs and one impeding vehicle, while the first CAV is the real Mazda CX-7 embedded in it.

*1) Hardware and Localization Setup:* In this test, one real internal combustion engine vehicle is involved. The vehicle actuation control is built based on two servo motors paired with acceleration and steering control robots as shown in Figure 7. The vehicle localization depends on Novatel PwrPak 7 GPS-RTK paired with an inertial navigation system (INS). INS can produce vehicle speed $v_t$, acceleration $a_t$, and orientation $\theta_t$ measurements while GPS-RTK can provide vehicle position $(x_t, y_t)$ measurements. Also, an onboard stereo camera is installed inside the vehicle to capture the mixed-reality simulation stream. The communication is implemented using the ROS platform. A properly tuned Kalman Filter is implemented to fuse the speed, acceleration, and position information for a more accurate and efficient estimation of vehicle position, velocity, and acceleration.

*2) Low-Level Vehicle Control:* In our experiment platform, we implement a specific control configuration to control the real ego vehicle as shown in Fig.7. During our experiment, the high-level planner was updated at 10 Hz, which cannot be used directly for control command outputs due to low frequency. Therefore, instead of using the predicted planner trajectories as references for Mazda's low-level MPC controller, we used a lane change logic flow to select reference trajectories for Mazda's low-level MPC controller to pair with lane change decisions made by the high-level planner. Our lane change logic aims to detect the imminent lane change requirement from the high-level planner, and then implements the lane change task based on a low-level lane change trajectory planner if a lane change is required within a certain number of steps $k$. The logic flow is shown in Figure 8. Considering the total number of steps from the high-level planner is $T$, then if the standard deviation of the first $k$ lane command is higher than a predefined value $\epsilon_l$, the lane change is considered to be required from the planner. Once the lane change is required, the low-level controller chooses the endpoint of the lane change task from the high-level planner $P_{end}(s_k, l_k)$. The other endpoint will be the current ego vehicle poses $P_{end}(s_0, l_0)$ as shown in Figure 9 In our experiment, to guarantee the smoothness of the lane change action, a trajectory generator is implemented to add round corners to the straight line between two endpoints. This results in smoother reference poses for the low-level MPC to track.

A Nonlinear MPC (NMPC) is implemented for reference tracking. An Euler discretized kinematic bicycle model with some minor modifications is utilized in the NMPC. The time interval is set to be 0.5 s with an 11-step horizon length to construct the MPC optimization problem to balance the computation efficiency and tracking accuracy [31]. We use the ACADO toolkit [32] to deploy the NMPC. The control inputs are acceleration $a_i$ and steering rate $\delta_i$. We use the vector $u_i = [a_i, \delta_i]^T$ to represent the control inputs. The state vector in the vehicle model is $X_i = [x_i, y_i, \phi_t, v_i, \theta]^T$ where $[x_i, y_i]$ represent the 2D position of the vehicle, $\phi_i$ represents the heading of the vehicle, $\theta_i$ represents the steering angle and $v_i$ represents the vehicle speed. Here we use the function $X_{i+1} = f(X_i, u_i)$ to represent the discrete form of the vehicle model. When formulating the MPC problem, the constraints are applied to the steering angle based on the vehicle parameters. The formulation is shown below.

$$\begin{aligned}
\min_{u_0,\dots,u_{T-1}} \quad & \sum_{i=0}^{T-1}[\|X_i - X_i^{ref}\|_Q^2 + \|u_i\|_R^2] \\
& + \|X_T - X_T^{ref}\|_{Q_T}^2 + q_{\epsilon,a}\epsilon_a^2 + q_{\epsilon,\theta}\epsilon_\theta^2 & (18a) \\
\text{s.t.} \quad & X_{i+1} = f(X_i, u_i) & (18b) \\
& \theta_i \in [\theta_{min} - \epsilon_\theta,\ \theta_{max} + \epsilon_\theta] & (18c) \\
& a_i \in [a_{min} - \epsilon_a,\ a_{max} + \epsilon_a] & (18d) \\
& \delta_i \in [\delta_{min},\ \delta_{max}] & (18e) \\
& \epsilon_\theta, \epsilon_a \geq 0 & (18f)
\end{aligned}$$

where $X^{ref}$ represents the constructed reference trajectory. We also incorporate slack variables for steering angle $\epsilon_\theta$, and acceleration $\epsilon_a$ in the constraints setup.

The high-level MIQP MPC is run on a Laptop with Ubuntu 20.04 OS, Intel(R) Core(TM) i7 (4.8 GHz) processor and 16 GB RAM while the NMPC is run on a Laptop with Ubuntu 20.04 OS, Intel(R) Core(TM) i9 (3.7 GHz) processor and 64 GB RAM. The communication is set up via ROS.

*3) Results:* In ViL experiments, we focus on the performance of the ego vehicle. Due to safety and logistical reasons, other three CAVs and the impeding vehicle are run in simulation only as "ghost vehicles". To record vehicle specific fuel consumption data, diagnostics data is collected through the OBD-II port of the CX-7. Fuel consumption is calculated from the mass air-flow (MAF) and air-fuel ratio measurements [26]. OBD-II logging is done by serially requesting and receiving data through an ELM327 device by utilizing the `python-OBD` library.

Rate of fuel consumption is calculated as $\dot{m}_f = \text{MAF}\frac{\lambda_c}{\lambda}$ where, $\lambda_c$ denotes the air-fuel ratio and $\lambda = 14.64$ denotes stoichiometric air-fuel ratio. However, when the engine is not generating propulsive torques, as in coasting or braking, the ECU switches to decelerating fuel cutoff (DFCO) for fuel saving. The MAF readings may still be non-zero even though the fuel consumption is zero due to cutoff. To handle this, the fuel rate is set to zero when the engine load torque $\tau_l$ is below coasting torque $\tau_c$ at speeds above 10 mph.

$$\dot{m}_f = 0 \leftarrow (\tau_l \leq \tau_c \wedge v > 10\,\text{mph}) \quad (19)$$

Table II shows the percent improvement in energy consumption (via fuel consumption) and travel time with GNEP-MPC over the baseline MPC in ViL experiments. The ViL results broadly align with the SiL results and up to 26.8 % (11.2 – 15.5 % on average) energy improvements were achieved with the GNEP-MPC compared to the baseline. Large fuel savings occur in the first and third sub-scenarios for the ego vehicle.

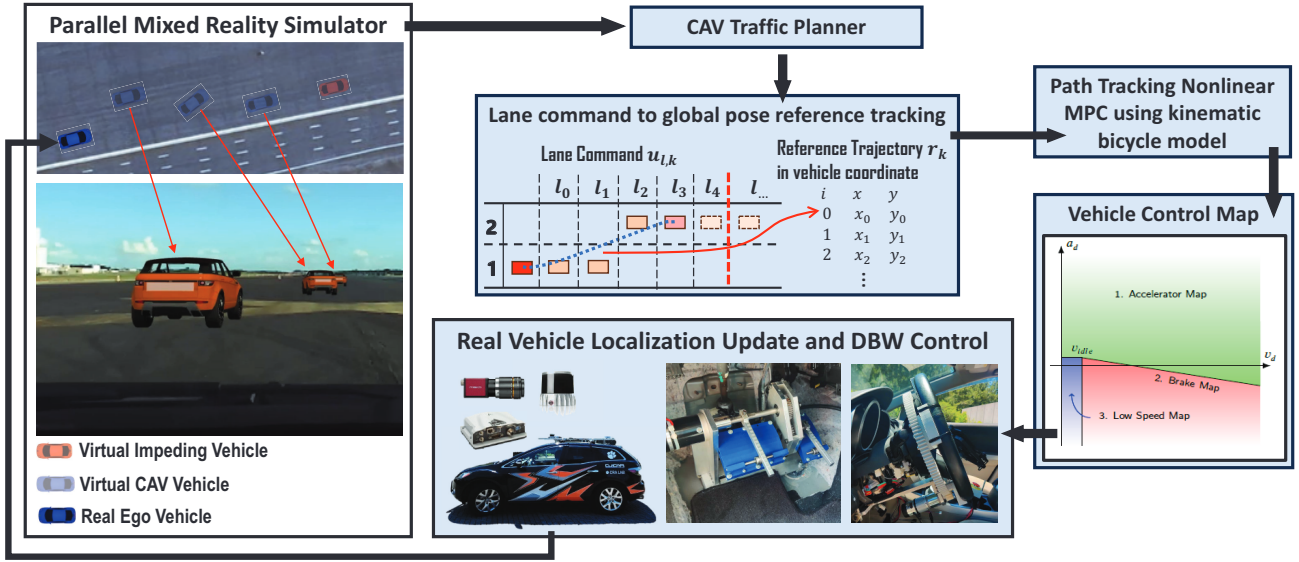

Fig. 7: Illustration of the experimental platform.

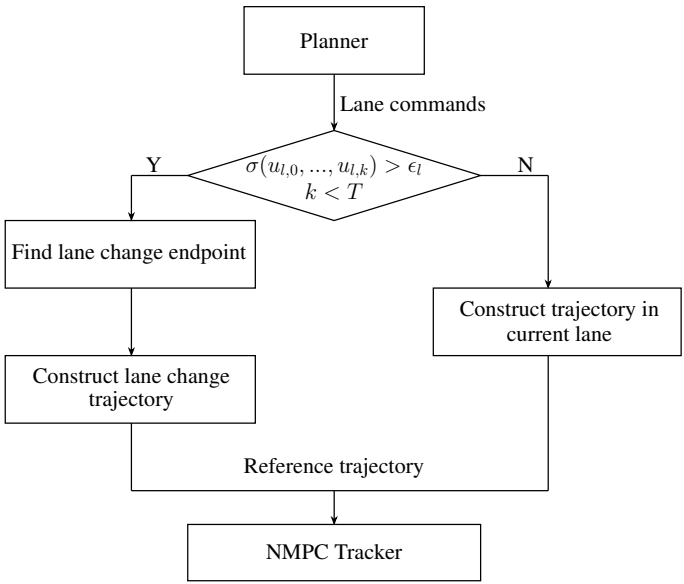

Fig. 8: Lane change logic for low-level MPC reference selection.

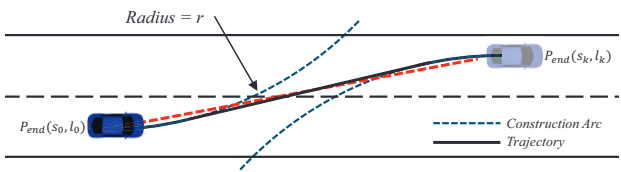

Fig. 9: Lane change trajectory generation demonstration.

In the fourth sub-scenario, lane change from right to left lane at 15 m/s could not be tracked fast enough by our low-level controller. The vehicle applied brakes to avoid collision with NV on the right lane due to insufficient lateral movement, thereby increasing its energy consumption. With re-tuning of the planner and tracker parameters, up to 4.7 % improvements could be achieved with slight a (0.4 %) travel time increase.

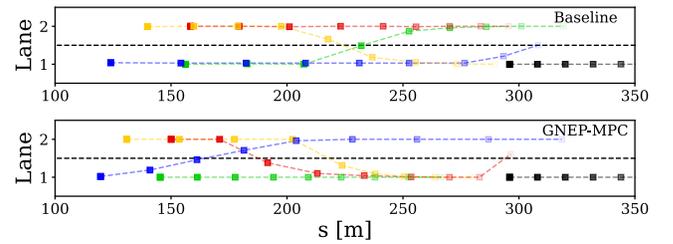

Fig. 10: ViL trajectory comparison: the real vehicle (Mazda CX-7) trajectory is shown in blue. Impeding vehicle is shown in black while rest are virtual CAVs. The points become lighter as time progresses. GNE approaching solutions reduce slow down behind the impeding vehicle by timely lane change. Sample trajectories from ViL experiments can be found at: [https://sites.google.com/g.clemson.edu/gnep-mpc-vil/home].

TABLE II: Percent improvement in energy consumption and travel time of real ego vehicle in ViL

| S.No. | Sub-scenario | % Energy | % Travel Time |
|---|---|---|---|
| 1 | (17, 14, 11, 8) | 26.8 | 4.0 |
| 2 | (11, 14, 17, 8) | 4.2 | 2.8 |
| 3 | (14, 11, 17, 8) | 26.2 | 8.0 |
| 4 | (17, 14, 8, 11) | -12.2/+4.7 | -3.9/-0.4 |

The fuel savings are a consequence of lower MPC costs attained due to approaching (open-loop) GNE solutions. As encoded in the MPC cost, low costs represent low vehicle accelerations (and closer reference velocity tracking) which implies less torque demand on the engine. A situation highlighting the advantage of GNEP-MPC is visualized in Figure 10 where the ego vehicle's reference velocity was 17 m/s. With the baseline planner, it had to slow down behind IV due to congestion. With the GNEP-MPC, congestion was prevented, resulting in reduced braking and consequently, lower accelerations. The closed-loop costs of the baseline and GNEP-MPC planners are compared for this sub-scenario in Figure 11. Trajectories from time step 40 to 85 are shown in Figure 10 and the reduced congestion results in lower corresponding costs in Figure 11.

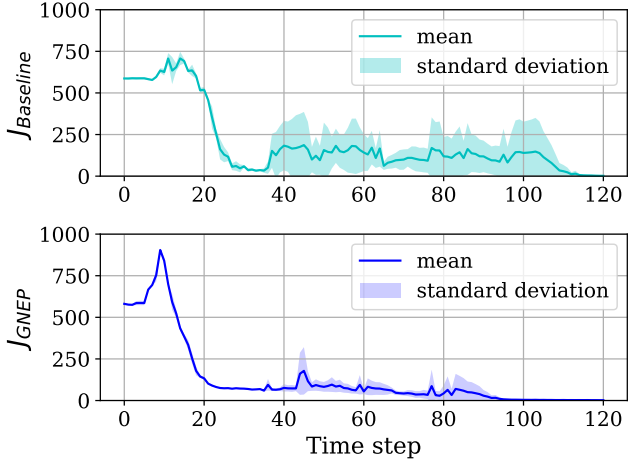

Fig. 11: ViL closed-loop cost comparison for sub-scenario 1.

## IV. CONCLUSION

In this work, we showed that V2V intention sharing in CAVs enables solution of an otherwise challenging multi-agent planning game as a distributed OCP for each CAV. In particular, this multi-agent planning method was implemented via distributed MPC by each CAV in lane change maneuvers. The MPC was deployed on a real vehicle and ViL experiments corroborated the energy and travel time benefits of approaching the (generalized) Nash equilibrium solution of the game, compared to unilateral prediction-based planning. On average, 11.2 – 15.5 % fuel savings were achieved. For ViL experiments, the real vehicle was embedded in a mixed reality testing environment involving simulated interacting agents localized in a virtual model of the ITIC test track in Greenville, SC, USA.

## APPENDIX A
## SiL LOW-LEVEL TRACKER

The following kinematic model $\dot{X} = f(X, U)$ of vehicle with states $X = [x, y, v, a, \theta]^T$ and controls $U = [u_a, u_\delta]^T$ is utilized in SiL tracker:

$$\dot{x} = v\cos(\theta + \beta) \quad (20a)$$
$$\dot{y} = v\sin(\theta + \beta) \quad (20b)$$
$$\dot{v} = a \quad (20c)$$
$$\dot{a} = (u_a - a)/\tau \quad (20d)$$
$$\dot{\theta} = (v/L_m)\sin\beta \quad (20e)$$

where, $\beta = \arctan(\frac{L_m}{(L_p+L_m)}\tan u_\delta)$, $\tau$ is the first order lag time constant and $L_m = L_p = 0.5\,\text{L}$ where $L$ is the wheelbase length.

The applied longitudinal control $u_a$ is that obtained from the planner while the lateral (steering) control law is as follows.

$$u_\delta = \begin{cases} \delta = k_1\,\Phi_e + \arctan(k_2\,\frac{l_e}{(k_3\,v+k_4)}) \text{ if } \delta \in (\delta_{min}, \delta_{max}) \\ \delta_{max} \text{ if } \delta \geq \delta_{max} \\ \delta_{min} \text{ if } \delta \leq \delta_{min} \end{cases}$$
(21)

where $\Phi_e = (\phi - \theta)$ ($\phi$ is the road orientation angle), $l_e = (u_l - l)$ ($u_l$ is that obtained from the planner and $l$ is the lane position of vehicle in planner coordinates) and $k_{1,2,3,4}$ are tuning parameters.

In simulation, the dynamics are integrated using Runge-Kutta 4 method. Gaussian noise $W \sim \mathcal{N}(0, \Sigma)$ is added to the model at each step such that $X_{k+1} = f_k(X_k, U_k) + W$. Hence, dynamic model mismatch between the planner and vehicle, and noise aspects of implementation are modeled in the tracker.

## ACKNOWLEDGMENT

The submitted manuscript has been created by UChicago Argonne, LLC, Operator of Argonne National Laboratory ("Argonne"). Argonne, a U.S. Department of Energy Office of Science laboratory, is operated under Contract No. DE-AC02-06CH11357. This paper and the work described were sponsored by the U.S. Department of Energy (DOE) Vehicle Technologies Office (VTO) under the Energy Efficient Mobility Systems (EEMS) Program, with support from EERE managers Avi Mersky, Erin Boyd and Alexis Zubrow.

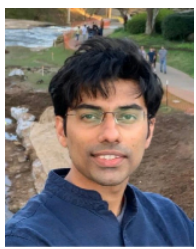
**Viranjan Bhattacharyya** is a PhD student in the Department of Mechanical Engineering at Clemson University, Clemson, USA. He received his MS in Mechanical and Aerospace Engineering from Illinois Institute of Technology, Chicago, USA. His research interest lies at the intersection of optimal control, game theory and motion planning with application in autonomous vehicles.

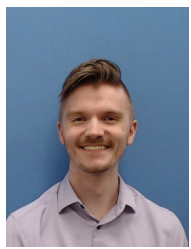
**Tyler Ard** Tyler Ard received the B.Sc. degree in mechanical engineering from Clemson University in 2017, and the Ph.D. degree in mechanical engineering from Clemson University in 2023. He is currently a postdoctoral appointee in the Vehicle and Mobility Simulation group at Argonne National Laboratory, where his research includes developing and verifying eco-driving strategies for ADAS and autonomous driving systems - in the aim of more efficient energy utilization of technologies already available today.

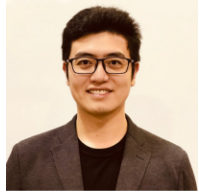
**Rongyao Wang** Rongyao Wang received the B.Sc degree in mechanical engineering from Clemson University in 2019, and the M.Sc degree in Mechanical engineering from Clemson University in 2021. After graduation, he worked in Stellantis, North America as a simulation software developer for a year. Currently, he is pursuing Ph.D degree in Automotive Engineering at Clemson University, where his research includes mixed reality based validation development for human-vehicle-in-the-loop research, aiming to achieve safer and more efficient collaborative intelligent traffic for the future.

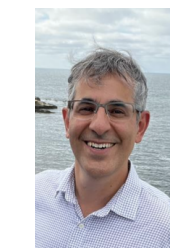
**Ardalan Vahidi** received the BSc and MSc degrees in civil engineering from the Sharif University of Technology, Tehran, Iran, in 1996 and 1998, respectively, the MSc degree in transportation safety from George Washington University in 2002, and the PhD degree in mechanical engineering from the University of Michigan, Ann Arbor in 2005. From 2012 to 2013, he was a Visiting Scholar with the University of California at Berkeley. He has held scientific visiting positions at the BMW Technology Office, Mountain View, California and IFP Energies Nouvelles, Rueil-Malmaison, France. He is currently Professor of Mechanical Engineering with Clemson University. His recent publications span topics connected and autonomous vehicles, efficient transportation, and human performance. He is a Fellow of IEEE and ASME.

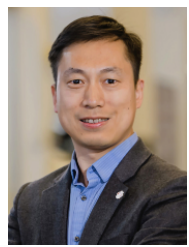
**Yunyi Jia** Yunyi Jia received the B.S. degree in automation from National University of Defense Technology, Changsha, China, in 2005, the M.S. degree in control theory and control engineering from South China University of Technology, Guangzhou, China, in 2008, and the Ph.D. degree in electrical engineering from Michigan State University, Michigan, USA, in 2014. He is currently the director of Collaborative Robotics and Automation Lab and an assistant professor in the Department of Automotive Engineering at Clemson University in Greenville SC USA. His research interests include robotics, automated vehicles and advanced sensing.

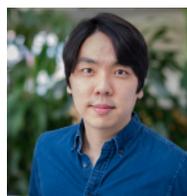
**Jihun Han** Jihun Han received his B.Sc., M.Sc., and Ph.D. in Mechanical Engineering from Korea Advanced Institute of Science and Technology (KAIST), South Korea, in 2009, 2011 and 2016, respectively. He was a postdoctoral research associate at IFP Energies Nouvelles, France, in 2016-2017, and at Oak Ridge National Laboratory, USA, in 2017-2018. He is currently a principal research engineer at the Vehicle and Mobility Simulation group and has worked at Argonne National Laboratory since 2018. His research interests include modeling, control, and simulation with an emphasis in intelligent transportation systems, and connected and automated vehicle systems.